\documentclass[11pt]{article}
\PassOptionsToPackage{authoryear}{natbib}
\usepackage[final]{neurips_2024}
\usepackage{hyperref} 
\usepackage{authblk} 
\usepackage{booktabs}
\usepackage{graphicx}
\usepackage{amsfonts}
\usepackage{multicol}
\usepackage{tabularx}
\usepackage{enumitem,amsmath,bm,amssymb}
\usepackage{pifont,makecell}
\usepackage{tikz}
\usepackage{CJKutf8}
\newcommand*\circled[1]{\tikz[baseline=(char.base)]{
		\node[shape=circle,draw,inner sep=0.5pt] (char) {#1};}}

\newcommand{\CHECK}{\textcolor{green}{\ding{51}}} 
\newcommand{\CROSS}{\textcolor{red}{\ding{55}}} 

\title{FunAudioLLM: Voice Understanding and Generation Foundation Models for Natural Interaction Between Humans and LLMs}

\author{Tongyi SpeechTeam}

\affil{Alibaba Group}
\affil{FunAudioLLM@list.alibaba-inc.com}
\date{} 
\begin{document}

\maketitle

\begin{abstract}
This report introduces FunAudioLLM, a model family designed to enhance natural voice interactions between humans and large language models (LLMs). At its core are two innovative models: SenseVoice, which handles multilingual speech recognition, emotion recognition, and audio event detection; and CosyVoice, which facilitates natural speech generation with control over multiple languages, timbre, speaking style, and speaker identity. SenseVoice-Small delivers exceptionally low-latency ASR for 5 languages, and SenseVoice-Large supports high-precision ASR for over 50 languages, while CosyVoice excels in multi-lingual voice generation, zero-shot in-context learning, cross-lingual voice cloning, and instruction-following capabilities. The models related to SenseVoice and CosyVoice have been open-sourced on Modelscope and Huggingface, along with the corresponding training, inference, and fine-tuning codes released on GitHub. By integrating these models with LLMs, FunAudioLLM enables applications such as speech-to-speech translation, emotional voice chat, interactive podcasts, and expressive audiobook narration, thereby pushing the boundaries of voice interaction technology. Demos are available at \url{https://fun-audio-llm.github.io}, and the code can be accessed at \url{https://github.com/FunAudioLLM}.

\end{abstract}

\section{Introduction}
In recent years, the advancement in artificial intelligence (AI) has dramatically transformed how humans interact with machines, such as GPT-4o \citep{DBLP:journals/corr/abs-2303-08774} and Gemini-1.5 \citep{DBLP:journals/corr/abs-2403-05530} and so on \citep{DBLP:journals/corr/abs-2308-12966,qwenaudio}. This transformation is particularly evident in the realm of voice processing, where capabilities such as high-precision speech recognition \citep{DBLP:conf/icml/RadfordKXBMS23}, emotion recognition \citep{DBLP:journals/corr/abs-2312-15185}, and voice generation \citep{DBLP:journals/corr/abs-2301-02111,cosyvoice} are paving the way for more intuitive and human-like interactions. In this report, we introduce FunAudioLLM, an innovative framework designed to facilitate natural voice interactions between humans and large language models (LLMs) \citep{qwen,DBLP:journals/corr/abs-2309-16609,DBLP:journals/corr/abs-2302-13971}. At the core of FunAudioLLM are our two groundbreaking models: SenseVoice, for voice understanding, and CosyVoice, for voice generation.

SenseVoice is our state-of-the-art voice understanding model, which excels in multiple domains of voice processing. We offer both SenseVoice-Small and SenseVoice-Large variants. We have open-sourced SenseVoice-Small, which supports multilingual recognition in Chinese, English, Cantonese, Japanese, and Korean, delivering extremely low inference latency by employing a non-autoregressive end-to-end architecture. This design choice results in a performance that is more than 5 times faster than Whisper-small and more than 15 times faster than Whisper-large \citep{DBLP:conf/icml/RadfordKXBMS23}. On the other hand, SenseVoice-Large supports speech recognition in over 50 languages, with significant advantages in recognizing Chinese and Cantonese. In addition to speech recognition, SenseVoice offers state-of-the-art capabilities in emotion recognition and audio event detection \citep{DBLP:journals/spm/MesarosHVP21}, making it an ideal choice for creating low-latency, human-like voice interaction systems.

Our suite of applications is further enriched by CosyVoice \citep{cosyvoice}, a family of fundamental speech generation models designed to produce natural-sounding voices for a variety of contexts. CosyVoice excels in generating multi-lingual voices tailored to specific speakers, zero-shot adaptation to new speakers \citep{DBLP:journals/corr/abs-2301-02111}, cross-lingual voice cloning \citep{DBLP:journals/corr/abs-2303-03926}, creating emotionally resonant voices \citep{DBLP:conf/interspeech/ShinLJHK22}, and offering nuanced control over speech output through instructional text \citep{ji2024controlspeechsimultaneouszeroshotspeaker}. CosyVoice supports five languages: Chinese, English, Japanese, Cantonese, and Korean. 
CosyVoice comes in three open-source models: CosyVoice-base-300M, which specializes in accurately representing speaker identity, zero-shot learning, and cross-lingual voice cloning; CosyVoice-instruct-300M, which focuses on generating emotionally expressive voices and allows for meticulous adjustments via instructional text, extending its capabilities to controllability over various aspects such as speaker identity \citep{DBLP:journals/corr/abs-2309-08140}, speaking style \citep{ji2024controlspeechsimultaneouszeroshotspeaker}, and fine-grained paralinguistic features \citep{DBLP:journals/corr/abs-2402-07383}; and CosyVoice-sft-300M, which has been fine-tuned on seven multilingual speakers and is ready for immediate deployment.

By integrating SenseVoice, CosyVoice, and LLMs like Qwen \citep{qwen}, FunAudioLLM offers a range of rich application demos. These include Speech-to-Speech Translation \citep{DBLP:conf/icassp/BerardBKP18}, which allows users to speak in foreign languages using their own voice; Emotional Voice Chat \citep{DBLP:journals/corr/abs-2401-00475}, which enables the model to understand and respond to emotions for more human-like interactions; Interactive Podcast \citep{DBLP:conf/iui/LabanYKCH22}, wherein users can engage in live discussions with multiple large models; 
and AudioBook \citep{DBLP:conf/lrec/ChalamandarisTKR14}, allowing the model to perform expressive, multi-character narration for audiobooks.

Overall, FunAudioLLM leverages the strengths of SenseVoice and CosyVoice to push the boundaries of voice interaction technology, enabling more natural and seamless communication between humans and large language models.

\section{FunAudioLLM Models}
\subsection{Overview of FunAudioLLM}
\begin{figure}[t!]
	\centering
	\includegraphics[width=\textwidth]{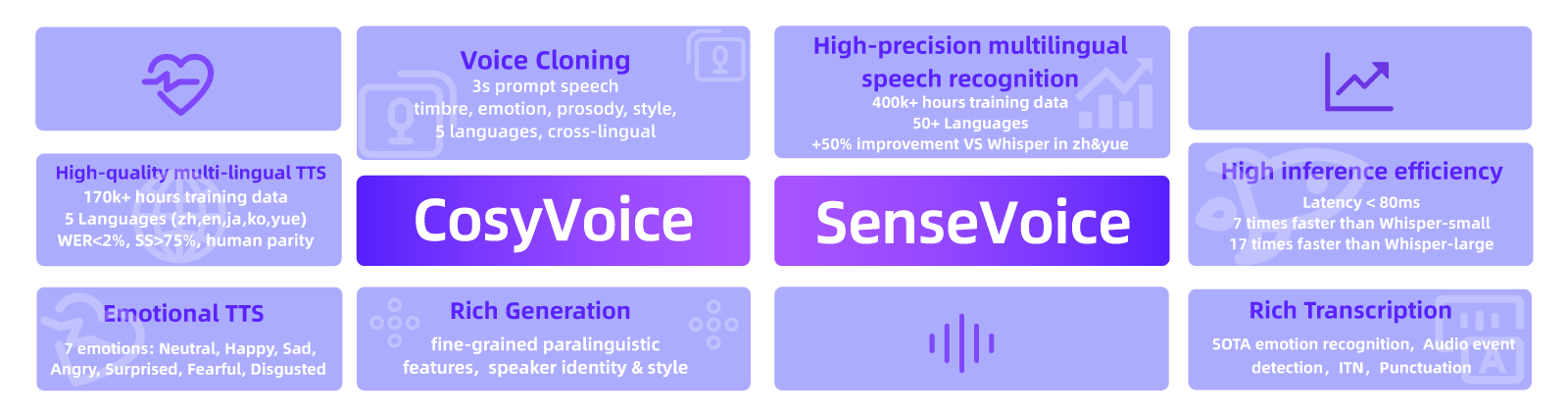}
	\caption{An overview of our FunAudioLLM models for voice understanding and generation.} 
	\label{fig:overview-funaudiollm}
\end{figure}
FunAudioLLM consists of two foundation models for voice understanding and generation, named SenseVoice and CosyVoice, respectively. SenseVoice supports multi-lingual speech recognition, which is trained on over 300k hours. Specifically, SenseVoice-Small is efficient in inference, in which the recognition latency is less than 80ms and is more than 5 and 15 times faster than Whisper-Small and Whisper-large, respectively, and SenseVoice-Large supports high-precision ASR for over 50 languages. Furthermore, SenseVoice supports rich transcription, including state-of-the-art emotion recognition, audio event detection, inverse text normalization \citep{DBLP:conf/interspeech/PusateriABPMN17} and punctuation \citep{DBLP:conf/icassp/ChenCLW20}.

Our voice generation model, CosyVoice, can generate multi-lingual speeches, which is trained on over 170k hours and five languages, including Chinese (ZH), English (EN), Japanese (JP), Cantonese (Yue) and Korean (KO). CosyVoice generated samples can achieve a WER of less 2\% and speaker similarity of over 75\%, which achieves the quality level of human parity.
CosyVoice supports zero-shot in-context learning, which enables voice cloning with a prompt speech of even 3 seconds. The timbre, emotion, prosody and style can be reproduced within or cross languages. We also released an instruction model, which can control speaker identity, speaking style (e.g., emotion) and other fine-grained paralinguistic features with natural textural instructions. An overview of FunAudioLLM models is shown in Figure \ref{fig:overview-funaudiollm}.

\subsection{Voice Understanding Model: SenseVoice}
\begin{figure}
\centering
\includegraphics[width=0.8\textwidth]{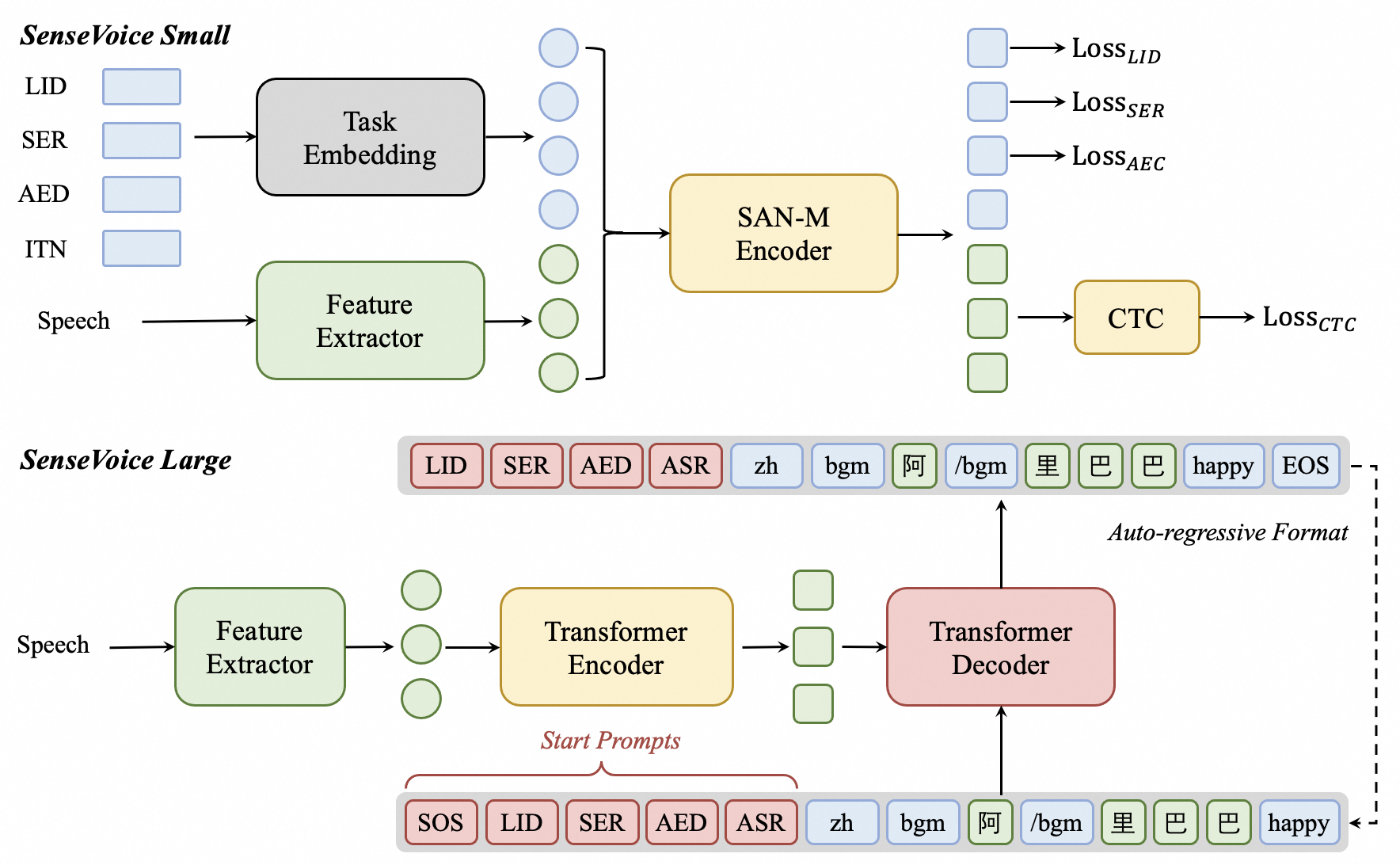}
\vspace{0.15cm}
\caption{SenseVoice is a comprehensive speech foundation model designed to perform various speech understanding tasks, including Automatic Speech Recognition (ASR), Language Identification (LID), Speech Emotion Recognition (SER), and Audio Event Detection (AED). SenseVoice-Small [Top]: An encoder-only model optimized for rapid speech understanding. It offers high-speed processing while supporting 5 languages. SenseVoice-Large [Bottom]: An encoder-decoder model aimed at achieving more precise speech understanding across a broader range of languages. It excels in accuracy and supports an extensive set of language capabilities.} \label{sensevoice}
\end{figure}
SenseVoice is a speech foundation model with multiple voice understanding capabilities, including automatic speech recognition (ASR), spoken language identification (LID), speech emotion recognition (SER), and audio event classification (AEC) or audio event detection (AED). Two models with different sizes and architectures are proposed to suit different requirements: SenseVoice-Small, an encoder-only speech foundation model for rapid speech understanding, and SenseVoice-Large, an encoder-decoder~\citep{DBLP:conf/nips/VaswaniSPUJGKP17} speech foundation model for more accurate speech understanding with more languages supported, as illustrated in Figure~\ref{sensevoice}.

\textbf{SenseVoice-Small} is a non-autoregressive encoder-only model for multi-lingual multi-style ASR and multiple speech understanding tasks. Given the input waveform, we first compute the 80-dimensional log-mel filter-bank, and then stack consecutive frames and down-sample them by a factor of 6. The extracted feature is then mapped to the dimension $D$ of the encoder, denoted as ${\rm \bf X}_{\rm speech} \in \mathbb{R}^{T \times D}$, where $T$ is the length of the down-sampled feature. The encoder is implemented as a memory-equipped self-attention network (SAN-M)~\citep{DBLP:conf/interspeech/GaoZLM20}. To specify the task, we prepend four embeddings to the speech feature as the input to the encoder:
\begin{equation}
    {\rm \bf X} = {\rm concat}({\rm \bf e}_{\rm LID}, {\rm \bf e}_{\rm SER}, {\rm \bf e}_{\rm AEC}, {\rm \bf e}_{\rm ITN/NoITN}, {\rm X}_{\rm speech})
\end{equation}
\begin{equation}
    {\rm \bf P} = {\rm Softmax}({\rm Linear}_{D \to |V'|}({\rm Encoder}({\rm \bf X})))
\end{equation}
${\rm \bf X} \in \mathbb{R}^{(T+4) \times D}$ and ${\rm \bf P} \in \mathbb{R}^{(T+4) \times |V'|}$. $V'$ is the vocabulary including tokens for ASR and other tasks. ${\rm \bf e}_{\rm LID}$, ${\rm \bf e}_{\rm SER}$, ${\rm \bf e}_{\rm AEC}$, ${\rm \bf e}_{\rm ITN/NoITN}$ are embeddings of four special tokens:

\bm{$\langle {\rm LID} \rangle$} indicates the LID task. If {$\langle {\rm LID} \rangle$} is prepended, the model is trained to predict the language token, at the corresponding position of the output. 
In the training stage, we randomly replace {$\langle {\rm LID} \rangle$} with the ground truth language token according to probability 0.8 so that the model can either predict the language token, or be configured with a specified language token in the inference stage.

\bm{$\langle {\rm SER} \rangle$} indicates the SER task. If {$\langle {\rm SER} \rangle$} is prepended, the model is trained to predict the speech emotion label, at the corresponding position of the output.

\bm{$\langle {\rm AEC} \rangle$} indicates the AEC task.  If {$\langle {\rm AEC} \rangle$} is prepended, the model is trained to predict the audio event label,  at the corresponding position of the output.

\bm{$\langle {\rm ITN} \rangle$} or \bm{$\langle {\rm NoITN} \rangle$} specify the transcription style. If $\langle {\rm ITN} \rangle$ is provided, the model is trained to transcript with inverse text normalization (ITN) and punctuation. If {$\langle {\rm NoITN} \rangle$} is provided, the model is trained to transcript without ITN and punctuation. 

In the training stage, the LID, SER, and AEC tasks are optimized using the cross-entropy loss. The ASR task is optimized using the CTC loss~\citep{DBLP:conf/icml/GravesFGS06}.

\textbf{SenseVoice-Large} is an autoregressive encoder-decoder model for multi-lingual ASR and multiple speech understanding tasks. Similar to Whisper~\citep{DBLP:conf/icml/RadfordKXBMS23}, SenseVoice-Large specifies tasks by a sequence of input tokens to the decoder. Specifically, we specify whether to predict language, speech emotion, and audio events with timestamps by including {$\langle {\rm LID} \rangle$}, {$\langle {\rm SER} \rangle$}, {$\langle {\rm AED} \rangle$} tokens respectively. Compared to SenseVoice-Small, the advantage of SenseVoice-Large is the transcription accuracy and supporting for a vast number of languages (50+).

Table~\ref{tab:example_sensevoice} gives examples of transcriptions of Whisper, SenseVoice-S, SenseVoice-L, and the ground truth of the ASR task.
\begin{table}[h]
    \footnotesize
    \centering
    \scalebox{1.0}{
    \begin{tabularx}{\textwidth}{X}
        \toprule
        \small{\textbf{Whisper}}  \\
            Absolute shock, but in a great way. Wow. That was awesome. That was awesome. What way to open a song. That was awesome. Awesome. $\cdots$ \\
            \midrule
        \small{\textbf{SenseVoice-S}}  \\
            $<music>$ Absolute shocked but in a great way my. $<happy>$ That was awesome, that was awesome what way to open a song that was awesome, awesome, $\cdots$ \\
            \midrule
        \small{\textbf{SenseVoice-L}}  \\
            $<music>$ Absolutely shocked but in a great way. That was awesome, $<music>$ that was awesome $<happy>$ what way to open a song, that was awesome, awesome,  $\cdots$ \\
            \midrule
        \small{\textbf{Ground Truth}}  \\
            Absolutely shocked, but in a great way. Who am I? Wow.  That was awesome. That was awesome. What way to open a song. That was awesome. Awesome. $\cdots$ \\
        \bottomrule
    \end{tabularx}}
    \vspace{0.15cm}
    \caption{Examples of transcriptions of Whisper, SenseVoice-S, SenseVoice-L, and the ground truth.}
    \label{tab:example_sensevoice}
\end{table}

\subsection{Semantic Speech Tokenizer}

A speech tokenizer transforms vocal signals into discrete tokens, enabling their modeling and prediction by autoregressive transformers for speech generation.
Our preliminary experiments indicated that the choice of speech tokenizer is pivotal for overall system performance as well as the requirements of both data quality and volume.
We evaluated three classes of speech tokenizers: 1) those based on residual quantization like SoundStream \citep{DBLP:journals/taslp/ZeghidourLOST22}, Encodec \citep{defossez2022highfi} and FunCodec \citep{DBLP:journals/corr/abs-2309-07405}; 2) those utilizing multi-grouped quantization, such as HifiCodec \citep{DBLP:journals/corr/abs-2305-02765}; and 3) ``semantic'' speech tokens, specifically HuBERT\citep{DBLP:journals/taslp/HsuBTLSM21}. All the above tokenizers are trained in the unsupervised or self-supervised manners. 
Thus, their association to semantic content is often tenuous, contributing to an unstable synthesis process and a substantial demand for clean training data. 
Moreover, unsupervised tokenizers are susceptible to data noise, necessitating meticulously curated clean data sets.

\begin{figure*}[h]
	\centering
	\includegraphics[width=0.3\linewidth]{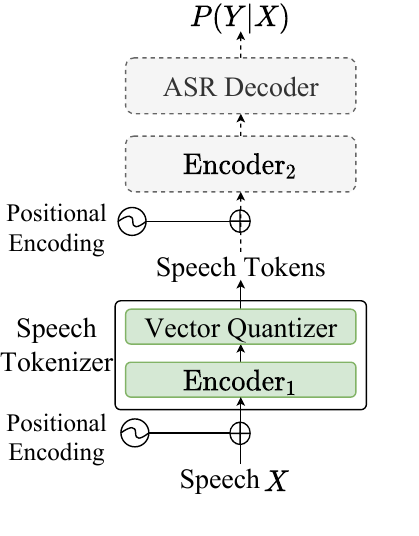}
	\vspace{-0.15cm}
	\caption{An illustration of our supervised semantic speech tokenizer.}
	\label{fig:speech-tokenizer}
\end{figure*}

Building on the success of SenseVoice models, we introduce a supervised semantic speech tokenizer, denoted as $\mathcal{S}^3$ \citep{cosyvoice}. 
Using the pre-trained SenseVoice-Large model as a foundation, we incorporate a vector quantizer subsequent to the encoder's initial six layers, delineated in Figure \ref{fig:speech-tokenizer}.
Importantly, the integration of an additional positional embedding post-quantization enhances temporal information.
The combination of $\textbf{Encoder}_1$ and vector quantizer is considered as the speech tokenizer, employing the index of the closest code vector as speech tokens.
The vector quantizer utilizes a solitary codebook with an expansive dictionary containing 4,096 entries.
The derived token sequence exhibits a frequency of 50 Hz, thereby reducing the computational load on text-to-token generation within language models.

Since the speech tokenizer is trained to minimize the recognition errors of rich text in an end-to-end manner, the extracted tokens have a strong semantic relationship to textual and paralinguistic information. Furthermore, our $\mathcal{S}^3$ tokenizer benefits from supervised training, enhancing its robustness to data noise and reducing the reliance on pristine data collection. Consequently, a broader spectrum of data can be utilized for training the model.

\subsection{Voice Generation Model: CosyVoice}
\begin{table}[t!]
	\centering
	\setlength\tabcolsep{3pt}
	\scalebox{0.95}{
		\begin{tabular}{lcccccc}
			\toprule
			Projects & Languages & Zero-shot & \makecell[c]{Style\&Speaker\\Control} & \makecell[c]{Fine-\\grained} & SFT & Server \\
			\midrule
			Bark & 13 & \CROSS & \CROSS & \CHECK & \CROSS & \CROSS \\
			\hline
			ChatTTS & en, zh & \CROSS & \CROSS & \CHECK & \CHECK & WebUI \\
			\hline
			parler-tts & en & \CROSS & \CHECK & \CROSS & \CHECK & \CROSS \\
			\hline
			EmotiVoice & en, zh & \CROSS & \CHECK & \CROSS & \CHECK & WebUI \\
			\hline
			GPT-SoVITS & en, zh, jp & \CHECK & \CROSS & \CROSS & \CHECK & WebUI \\
			\hline
			OpenVoice & \makecell[c]{en,sp,fr, zh,jp,kr} & \CHECK & \CHECK & \CHECK & \CROSS & \CROSS \\
			\hline
			CosyVoice & \makecell[c]{en, zh, jp, yue, kr} & \CHECK & \CHECK & \CHECK & \CHECK & \makecell[c]{WebUI, gRPC} \\
			\bottomrule
	\end{tabular}}
	\vspace{0.15cm}
	\caption{Comparison on released features between CosyVoice and other open-sourced projects.}
	\label{tab:feature-comp}
\end{table}
CosyVoice, a family of fundamental speech generation models \citep{cosyvoice}, utilizes $\mathcal{S}^3$ tokens to synthesize natural-sounding voices suitable for various applications.
As a versatile model, CosyVoice excels in tasks such as generating multi-lingual voices tailored to specific speakers, adapting to new speakers without training (zero-shot in-context learning), replicating voices across different languages (cross-lingual voice cloning), creating emotionally resonant voices, and offering nuanced influence over speech output through instructional text.
CosyVoice supports five languages, including Chinese (ZH), English (EN), Japanese (JP), Cantonese (Yue) and Korean (KO). We released three open-source models. The first, CosyVoice-base-300M, excels in accurately representing speaker identity, adapting to contexts without any finetuning, and cloning voices across languages. The second, CosyVoice-instruct-300M, is adept in generating emotionally expressive voices and allows for meticulous adjustments via instructional text. Lastly, CosyVoice-sft-300M has been fine-tuned on seven multi-lingual speakers and is ready for immediate deployment. All of them share the common model architecture and learning framework. Compared with other open-sourced projects, CosyVoice released a widest spectrum of supporting features as shown in Table \ref{tab:feature-comp}.

\begin{figure}[h]
	\centering
	\includegraphics[width=1.0\linewidth]{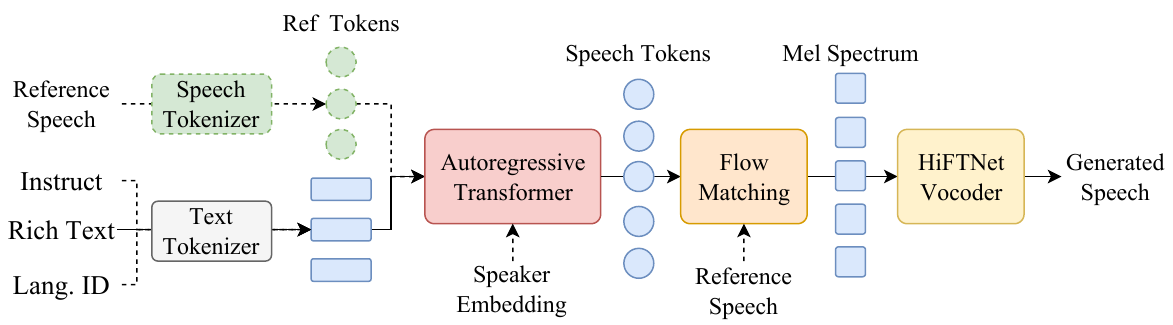}
	\vspace{-7mm}
	\caption{A semantic diagram of CosyVoice models.}
	\label{fig:cosyvoice}
\end{figure}
\subsubsection{System Overview}
CosyVoice incorporates an autoregressive Transformer-based language model (LM) to generate speech tokens for the input text.
An ordinary differential equation based (ODE-based) diffusion model, flow matching \citep{DBLP:conf/iclr/LipmanCBNL23}, reconstructs Mel spectrum from the generated tokens. Subsequently, a HiFTNet-based vocoder \citep{DBLP:journals/corr/abs-2309-09493} is followed to synthesize waveforms from the reconstructed Mel spectrum. Dashed models are optional for certain applications, such as cross-lingual cloning and speaker fine-tuned inference.

\subsubsection{Model Training}
At the training stage, the autoregressive language model (LM) is trained using a teacher-forcing paradigm.
In this process, tokenized text and a left-shifted version of the speech tokens are provided as input to predict the subsequent speech tokens.

The flow matching model is developed to estimate the conditional probabilities $P(S|X,v,S_{ref})$, where $X$ and $v$ denote the speech tokens and speaker embeddings \citep{DBLP:conf/interspeech/WangZCC023}, respectively. $S$ and $S_{ref}$ represent the Mel spectrum of target and reference speech, respectively. 
A convolutional Transformer U-Net \citep{DBLP:journals/corr/abs-2309-03199} is employed to ascertain the vector field between the prior distribution and the desired one, which is derived from the optimal transport ODE. The straightforward nature of resolving the OT-ODE allows for a significantly reduced number of iterations during the inference stage, typically only five to ten iterations are required to produce a satisfactory Mel spectrogram. We also employ the classifier-free guidance (CFG) \citep{DBLP:journals/corr/abs-2207-12598} technique and mask out the 70\%$\sim$100\% proceeding feature conditions to boost the in-context learning ability.

For the synthesis of waveforms from the predicted Mel spectrograms, we utilize a vocoder based on HiFTNet \citep{DBLP:journals/corr/abs-2309-09493}. Modifications have been made on HiFTNet to support streaming generation, including the replacement and redesign of certain components. Complete details regarding these adjustments are available in our released code.

\subsubsection{Zero-shot In-context Learning}
\begin{figure*}[t!]
	\centering
	\includegraphics[width=0.8\linewidth]{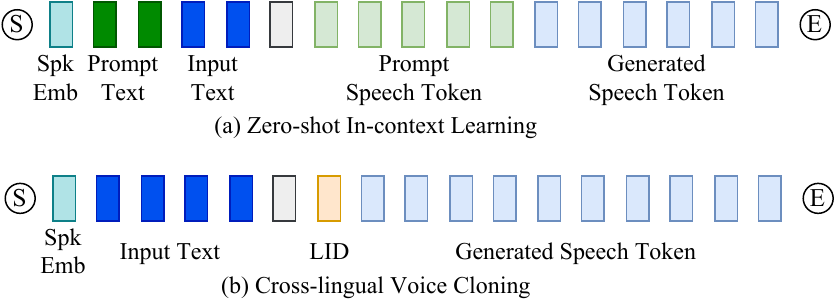}
	\caption{Sequence construction for (a) zero-shot in-context learning and (b) cross-lingual voice cloning. LID represents language identifier.}
	\label{fig:icl-seq}
\end{figure*}

CosyVoice models exhibit zero-shot in-context learning capabilities, allowing for the replication of an arbitrary voice with only a brief reference speech sample. This process entails the careful construction of input sequences for the token language model (LM), depicted in Figure \ref{fig:icl-seq}.
For prompt speech and input text in the same language, we merge them to form a unified input, treating the prompt speech tokens as pre-generated. With this input sequence, the autoregressive LM iteratively predicts subsequent tokens until it encounters the ``end of sequence'' token $\circled{E}$.
However, when the prompt speech and input text differ linguistically, we omit the text and tokens associated with the prompt to prevent prosodic characteristics of the original language from influencing the target language.
It is important to note that the prompt text, which corresponds to the prompt speech's content, can be transcribed either through human annotation or ASR models, such as SenseVoice. Similar to the prompt text, the prompt tokens are extracted from the prompt speech with $\mathcal{S}^3$ tokenizer.

After generating the speech tokens, they are appended after the prompt tokens, forming a composite condition for the flow-matching model. Additionally, the speaker embedding and the Mel spectrogram of the prompt speech are incorporated to further enhance timbre and environmental consistency.

\subsubsection{Instruction Fine-tuning}
To enable further controllability on CosyVoice, we experiment with integrating additional instruction fine-tuning \citep{DBLP:journals/corr/abs-2308-14430}. 
CosyVoice-instruct extends CosyVoice-base with enhanced instruction-following capabilities. Specifically, it supports controllability over various aspects such as speaker identity (i.e., speaker's characteristics), speaking style (including emotion, gender, speaking rate, and pitch), and fine-grained paralinguistic features. These features include the ability to insert laughter, breaths, speaking while laughing, and emphasizing certain words. Table \ref{tab:example_instruct} shows some examples of speaker identity, speaking style, and fine-grained paralinguistic features. 

\begin{table}[h]
    \footnotesize
    \centering
    \scalebox{1.0}{
    \begin{tabularx}{\textwidth}{X}
        \toprule
        \small{\textbf{Speaker Identity}}  \\
            1. Selene 'Moonshade', is a \textbf{mysterious}, \textbf{elegant dancer} with a connection to the night. Her movements are both \textbf{mesmerizing} and \textbf{deadly}.$<$endofprompt$>$Hope is a good thing.\\
            2. Theo 'Crimson', is a \textbf{fiery}, \textbf{passionate} rebel leader. Fights with fervor for justice, but struggles with \textbf{impulsiveness}.$<$endofprompt$>$You don't know about real loss. \\
            \midrule
          \small{\textbf{Speaking Style}}  \\
            1. A \textbf{happy} \textbf{girl} with \textbf{high tone} and \textbf{quick speech}.$<$endofprompt$>$The sun is shining brightly today. \\
            2. A \textbf{sad woman} with \textbf{normal tone} and \textbf{slow speaking speed}.$<$endofprompt$>$I failed my important exam. \\
            \midrule
        \small{\textbf{Fine-grained Paralinguistics}} \\
            1. Well that's kind of scary \textbf{[laughter]}. \\
            2. I don't think I over eat yeah \textbf{[breath]} and um I do exercise regularly. \\
            3. Well that pretty much covers \textbf{$<$laughter$>$the subject$<$/laughter$>$} well thanks for calling me. \\	
            4. The team's \textbf{$<$strong$>$unity$<$/strong$>$} and \textbf{$<$strong$>$resilience$<$/strong$>$} helped them win the championship. \\
        \bottomrule
    \end{tabularx}
    }
	\vspace{0.15cm}
    \caption{Examples of speaker identity, speaking style, and fine-grained paralinguistics. }
    \label{tab:example_instruct}
\end{table}



\section{Dataset}
\subsection{Training Set for SenseVoice}
\begin{figure}[ht]
	\centering
	\includegraphics[width=\textwidth , trim={0.2cm 0.2cm 0.2cm 1.0cm}, clip]{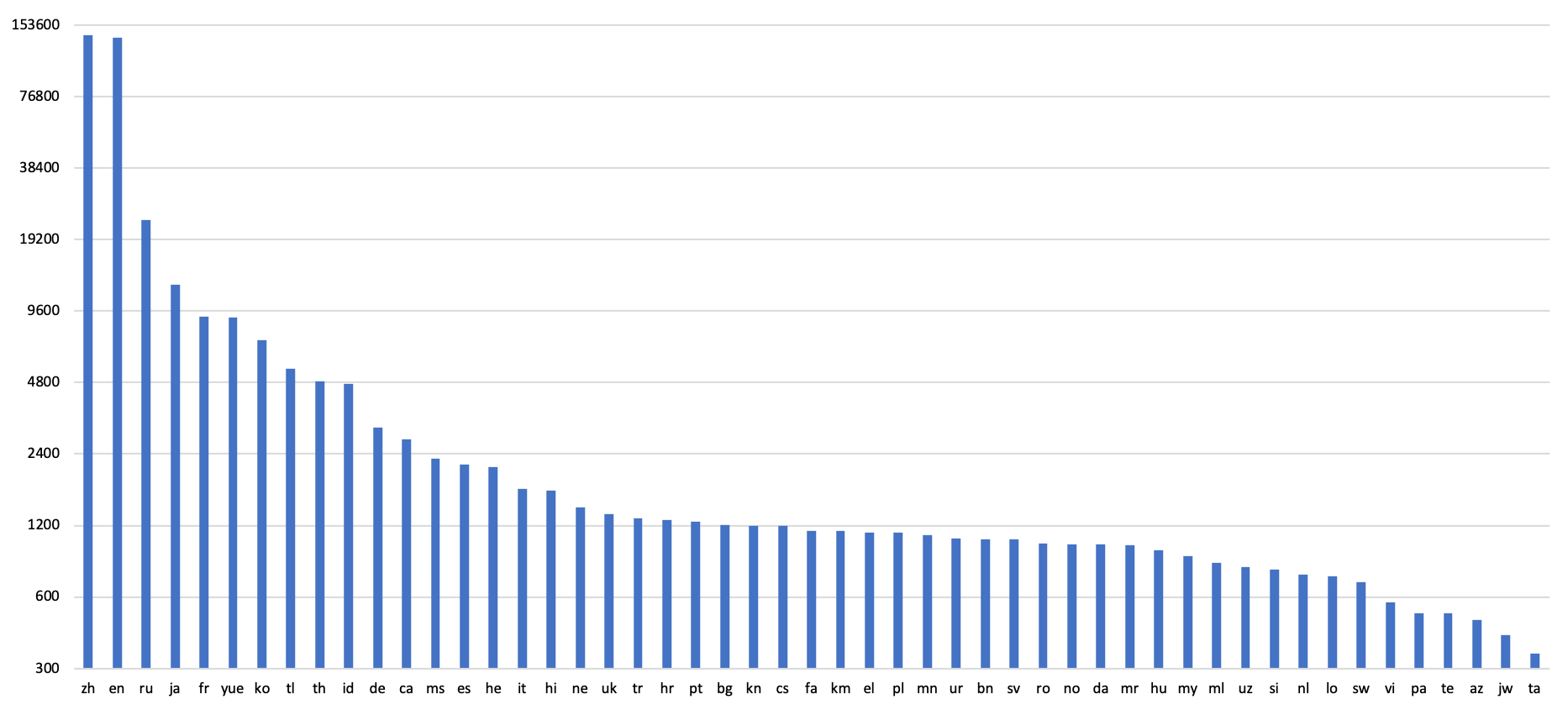}
	\vspace{0.15cm}
	\caption{Hours of SenseVoice training data across languages (in log scale).}
	\label{asr_dataset}
\end{figure}
Figure~\ref{asr_dataset} provides an overview of the dataset utilized for training the SenseVoice models. The SenseVoice-Small model was trained on an extensive audio data corpus of approximately 300,000 hours, covering 5 languages including Chinese, Cantonese, English, Japanese, and Korean. To further enhance the multilingual ability of SenseVoice-Large, an additional 100,000 hours of diverse multilingual data were integrated into the training corpus.
To obtain rich transcription labels from speech data, we leveraged open-source models for audio event detection (AED) \footnote{\url{https://github.com/qiuqiangkong/audioset_tagging_cnn/tree/master}} and speech emotion recognition (SER) \footnote{\url{https://modelscope.cn/models/iic/emotion2vec_plus_large}} to generate pseudo labels, yielding an extensive rich transcribe dataset. Specifically, the AED data amounted to 150 million entries, while the SER data comprised 30 million entries.

\begin{table}[ht]
	\centering
	\setlength\tabcolsep{15pt}
	\scalebox{1.0}{
		\begin{tabular}{lr}
			\toprule
			Language & Duration (hr) \\
			\midrule
			ZH & 130,000 \\
			EN & 30,000 \\
			Yue & 5,000 \\
			JP & 4,600 \\
			KO & 2,200 \\
			\bottomrule
	\end{tabular}}
	\vspace{0.15cm}
	\caption{Hours of CosyVoice training data across languages.}
	\label{tab:dataset}
\end{table}
\subsection{Training Set for CosyVoice}
To train the CosyVoice models, we have amassed a considerable dataset comprising multiple languages. Throughout the collection process, we utilize specialized in-house tools for speech detection, signal-to-noise ratio (SNR) estimation, speaker diarization, and separation. Subsequently, pseudo text labels are generated using SenseVoice-Large and Paraformer. These labels undergo a refinement process with the aid of force-alignment (FA) models, which helps eliminate low-quality data and enhances the accuracy of punctuation. A comprehensive breakdown of the training data's duration across various languages is presented in Table \ref{tab:dataset}. 

For the CosyVoice-instruct model, we fine-tuned CosyVoice-base using instruction training data without incorporating speaker embedding in the autoregressive language model. Table \ref{tab:dataset_instruct} presents the duration of the training data for different types of instructions.

\begin{table}[h]
	\centering
	\setlength\tabcolsep{15pt}
	\scalebox{1.0}{
		\begin{tabular}{lr}
			\toprule
			Type & Duration (hr) \\
			\midrule
			Speaker Identity & 101 \\
			Speaking Style & 407 \\
			Fine-grained Paralinguistics & 48 \\
			\bottomrule
	\end{tabular}}
	\vspace{0.15cm}
	\caption{Duration statistics of instruction training data by type.}
	\vspace{-0.6cm}
	\label{tab:dataset_instruct}
\end{table}

\section{Experimental Results}
\subsection{Multilingual Speech Recognition}
\textbf{Metrics.} We use Character Error Rate (CER) to evaluate the models in five languages: Chinese, Cantonese, Japanese, Korean, and Thai, and use the Word Error Rate (WER) for all other languages. Both the ground truth transcriptions and the recognition outputs are standardized using text normalization before the error rate calculation, in alignment with the methodology used by Whisper. All Chinese characters were converted into the simplified Chinese version, together with an additional text normalization pipeline\footnote{\url{https://github.com/speechio/chinese_text_normalization/blob/master/python/cn_tn.py}}.

Results in Table~\ref{tab:performance1} show the comparison of Whisper, SenseVoice and Paraformer~\citep{DBLP:conf/interspeech/GaoZ0Y22,gao2023funasr,shi2024seaco} on popular open speech recognition benchmark datasets, including AISHELL-1 \citep{bu2017aishell}, AISHELL-2 \citep{du2018aishell}, WenetSpeech \citep{zhang2022wenetspeech}, Librispeech~\citep{panayotov2015librispeech}, and Common Voice \citep{ardila2019common}.  It can be seen that SenseVoice-S and SenseVoice-L outperform their Whisper counterparts by a significant margin in most test sets except Librispeech. 

\begin{figure}
	\centering
	\includegraphics[width=\textwidth , trim={0.2cm 0.2cm 0.2cm 1.0cm}, clip]{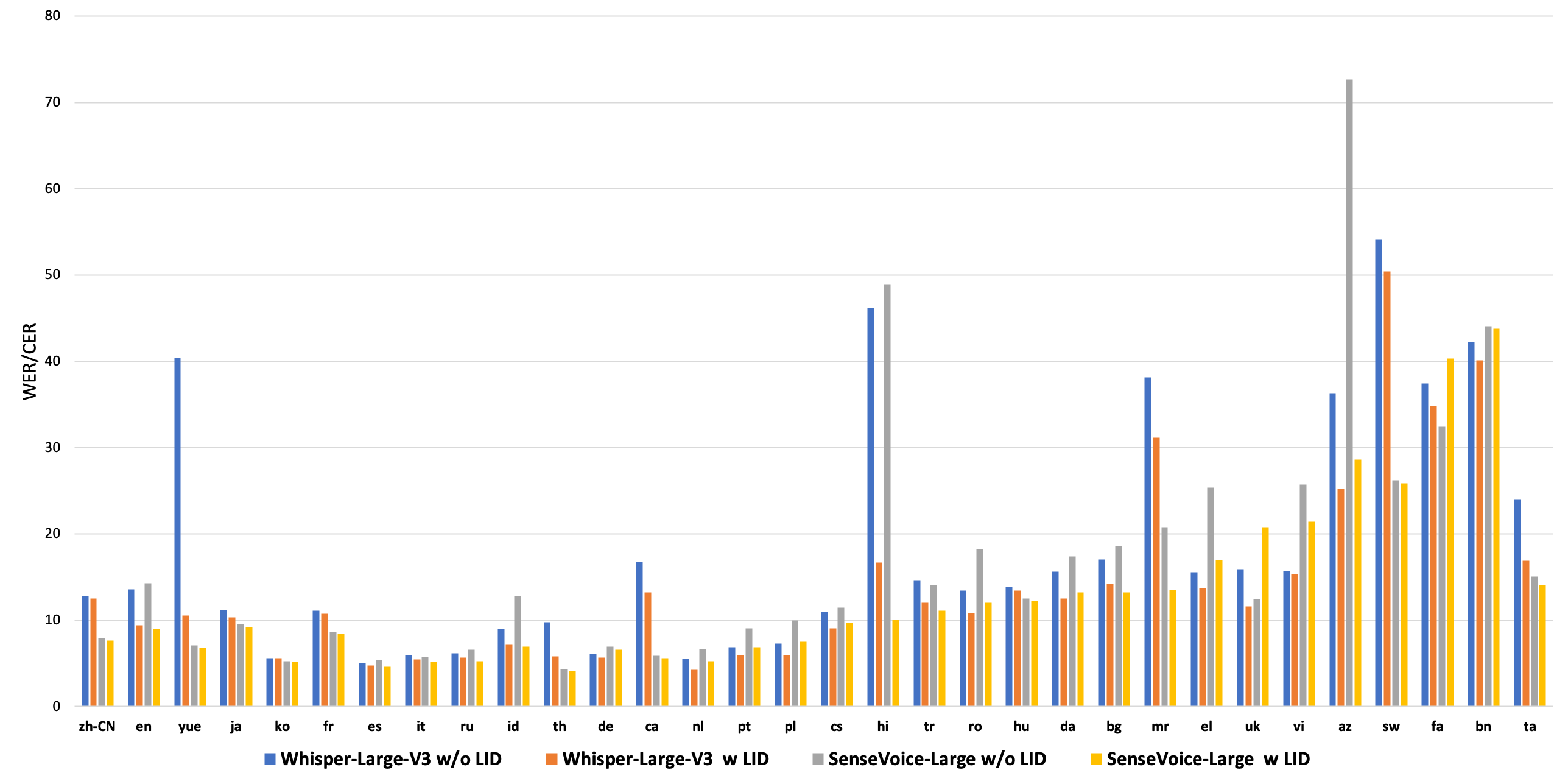}
	\vspace{0.15cm}
	\caption{ Comparison of SenseVoice and Whisper on Common Voice, with or without LID} \label{common_voice}
\end{figure}

Figure~\ref{common_voice} illustrates the comparative performance of SenseVoice-Large and Whisper-Large-V3 on a broader range of languages,  with or without ground truth LID as input. While SenseVoice-Large performs comparably with Whisper-Large-V3 in general, SenseVoice-Large obtains significantly better performance in languages like Cantonese (Yue), Catalan (CA), and Marathi (MR).

The evaluation of inference efficiency is shown in Table~\ref{tab:inference_efficiency}. The Real-time factor (RTF, the ratio of the transcribing time to the audio length) and 10s Audio Latency (the average time cost when transcribing a 10s audio.) are benchmarked on an A800 machine,  with a decoding batch size of 1. For the encoder-decoder-based model (Whipser-S, Whipser-L-V3, and SenseVoice-L), we perform beam search in decoding with a beam size of 5. Owing to its non-autoregressive architecture, SenseVoice-S obtains extremely low inference latency—more than 5 times faster compared to Whisper-small and more than 15 times faster compared to Whisper-L-V3. SenseVoice-L shows close performance with Whipser-L-V3.

\begin{table}[ht]
	
	\centering
	\scalebox{0.7}{
		\begin{tabular}{llccccc}
			\toprule
			& Whisper-S & Whisper-L-V3  & SenseVoice-S & SenseVoice-L  & Paraformer-zh \\
			\midrule
			AISHELL-1 test & 10.04 & 5.14  & 2.96 & 2.09 & \textbf{1.95}\\
			\midrule
			AISHELL-2 test\_ios & 8.78 & 4.96  & 3.80 & 3.04 & \textbf{2.85}\\
			\midrule
			WenetSpeech test\_meeting & 25.62 & 18.87  & 7.44 & \textbf{6.73} & 6.97 \\
			\midrule
			WenetSpeech test\_net & 16.66 & 10.48  & 7.84 & \textbf{6.01} & 6.74 \\
			\midrule
			LibriSpeech test\_clean & 3.13 & \textbf{1.82}  & 3.15 & 2.57 & - \\
			\midrule
			LibriSpeech test\_other & 7.37 & \textbf{3.50}  & 7.18 & 4.28 & -\\
			\midrule
			CommonVoice zh-CN  & 19.60  & 12.55 &  10.78 & \textbf{7.68} & 10.30\\
			\midrule
			CommonVoice en  & 14.85  & 9.39  & 14.71 & \textbf{9.00} & - \\
			\midrule
			CommonVoice yue & 38.97  & 10.41 & 7.09 & \textbf{6.78} & -  \\
			\midrule
			CommonVoice ja  & 19.51  & 10.34  & 11.96 & \textbf{9.19} & - \\
			\midrule
			CommonVoice ko  & 10.48  & 5.59  & 8.28 & \textbf{5.21} & - \\
			\midrule
			CommonVoice 5 lang. Average & 20.68 & 9.66 & 10.56 & \textbf{7.57} & - \\
			\bottomrule
	\end{tabular}}
	\vspace{0.15cm}
	\caption{Performance comparisons among different models on Chinese and English Open Corpus.}
	\label{tab:performance1}
\end{table}

\begin{table}[th]
	
	\centering
	\scalebox{0.8}{
		\begin{tabular}{llccccccccc}
			\toprule
			Model & Framework & Parameters & Support Language & RTF & 10s Audio Latency(ms)  \\
			\midrule
			Whisper-S & Autoregressive & 224M & 50+ & 0.042  & 518\\
			\midrule
			Whisper-L-V3 & Autoregressive & 1550M & 50+ & 0.111 & 1281 \\
			\midrule
			Paraformer-zh & Non-autoregressive & 220M & zh & 0.009 & 100\\
			\midrule
			SenseVoice-S & Non-autoregressive & 234M & zh,yue,en,ja,ko & 0.007 & 70 \\
			\midrule
			SenseVoice-L & Autoregressive & 1587M & 50+ & 0.110 & 1623 \\
			\bottomrule
	\end{tabular}}
	\vspace{0.15cm}
	\caption{Comparison of model architecture, parameter scale, supported languages, and inference efficiency of SenseVoice, Paraformer, and Whisper.}
	\label{tab:inference_efficiency}
\end{table}

\subsection{Speech Emotion Recognition}

\begin{table}[th]
	\centering
	\scalebox{0.75}{
		\begin{tabular}{l  ccc  c  c  cccc  cccc}
			\toprule
			& \multicolumn{3}{c}{EmoBox} & \multicolumn{1}{c}{Emo-Superb}  & \multicolumn{1}{c}{MerBench} & \multicolumn{4}{c}{SenseVoice-L} & \multicolumn{4}{c}{SenseVoice-S} \\
			Test set & UA & WA & F1 &  F1 & WF1 &  UA & WA & F1 & WF1 & UA & WA & F1 & WF1 \\
			\midrule
			CASIA  & 59.6&59.6&56.3  &--  &--  &  96.0&96.0&95.5&95.5   &70.0&70.0&70.3&70.3 \\
			CREMA-D & 76.8 & 76.5 &76.6   &67.7  &--&    90.1& 90.4&89.8&89.9    &73.1&74.0& 65.7&65.7 \\
			ESD     &84.6&84.6&84.3  &--  &--  &  93.2&93.2&92.2&92.2   &85.5&85.5&81.0&81.0 \\
			IEMOCAP & 73.5 & 72.9 & 73.1  & --   &69.7&  73.9 & 75.3 & 73.2&72.8   &70.5&65.7& 67.9&67.8 \\
			MELD    & 31.5 & 51.9 & 32.9  & --   &46.5&  58.7 & 63.1 & 50.9&65.7   &50.8&57.8& 44.6&57.7 \\
			MER2023 &61.2&65.2&62.3  &--  &67.5  &  70.9&69.2&55.6&57.4   &69.0&68.3&52.8&56.6 \\
			MSPPodcast  & 21.4 & 43.4 & 21.5  &38.4  &--  &  46.0 & 61.7 & 45.0&58.9   &49.4&64.1& 46.4&63.1 \\
			\bottomrule
	\end{tabular}}
	\vspace{0.15cm}
	\caption{SER performance comparisons on different evaluation benchmarks.}
	\label{tab:ser_performace}
\end{table}

We evaluate the SER ability of the SenseVoice on 7 popular emotion recognition datasets, including CREMA-D\citep{cremad}, MELD\citep{meld}, IEMOCAP\citep{IEMOCAPIE}, MSP-Podcast\citep{msppodcast}, CASIA\citep{casia}, MER2023\citep{lian2023mer2023multilabellearning} and ESD\citep{esd}. These corpora cover both Chinese and English, and scenarios like acts, TV dramas, and daily conversation. We report unweighted average accuracy (UA), weighted average accuracy (WA), macro F1 Score (F1), and weighted average F1 (WF1), and compare them with some recently published SER benchmarks (EmoBox \citep{emobox}, Emo-Superb\citep{emosuperb} and MerBench \citep{merbench}) from literature in Table \ref{tab:ser_performace}. We show that SenseVoice achieves a good performance on all test sets and all metrics even without fine-tuning on the target domain.

\begin{figure}[!h]
	\centering
	\includegraphics[width=0.7 \textwidth , trim={0.2cm 0.2cm 0.2cm 1.0cm}, clip]{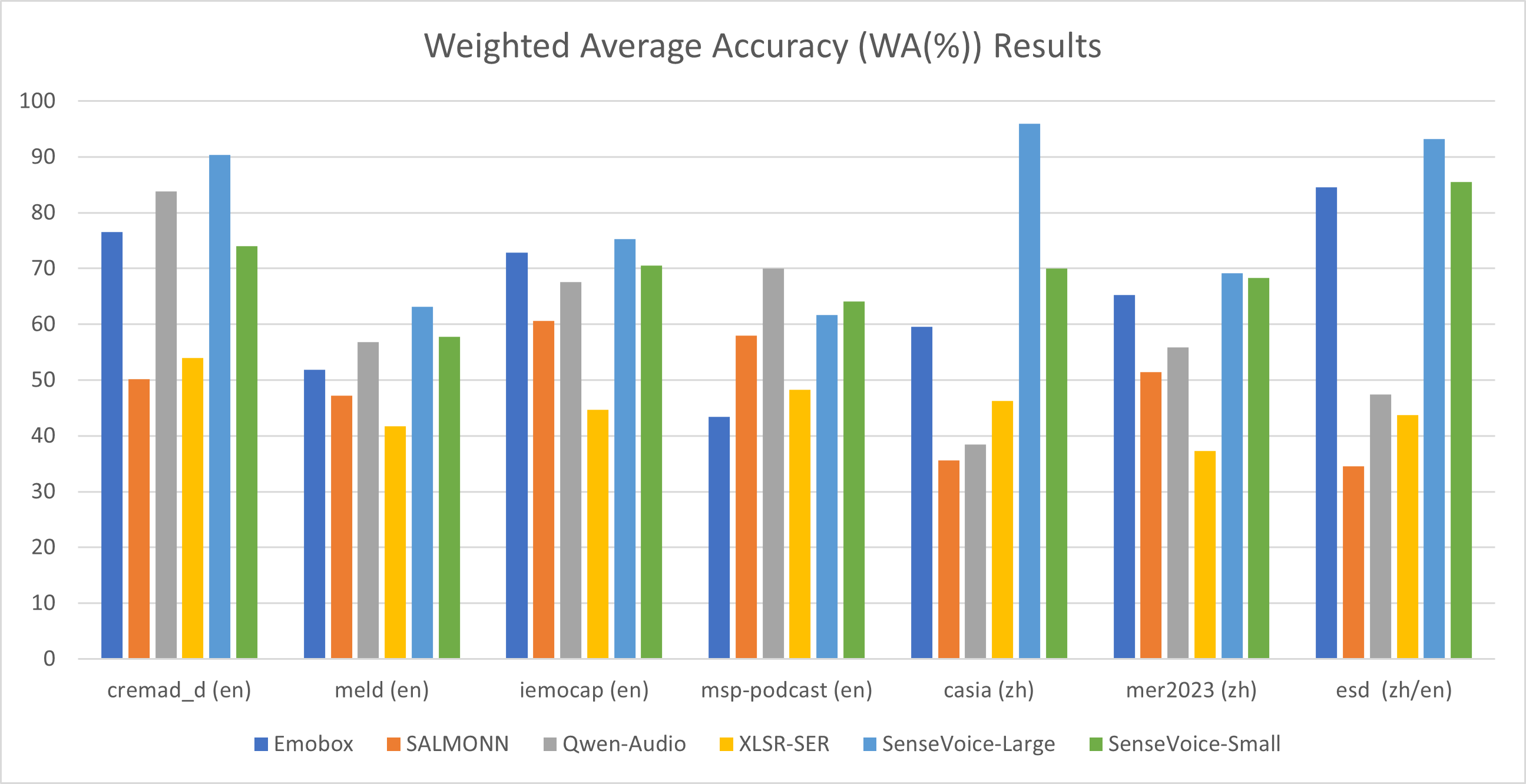}
	\vspace{0.15cm}
	\caption{Weighted Average Accuracy (WA(\%)) comparison with other open source SER models.} \label{ser_results}
\end{figure}

We further compare SenseVoice with some open-sourced SER models. Results are shown in Figure~\ref{ser_results}. XLSR-SER is the most popular SER model on HuggingFace\footnote{\url{https://huggingface.co/ehcalabres/wav2vec2-lg-xlsr-en-speech-emotion-recognition}}, and Qwen-Audio\citep{qwenaudio} and SALMONN\citep{salmonn} are two Audio-LLM models which can recognize speech emotion with natural language prompt. Results from EmoBox are also involved in the figure as references.
SenseVoice-Large achieves the best results on almost all datasets while the SenseVoice-Small also outperforms other baseline models on the majority datasets.

\subsection{Audio Event Detection}

Both SenseVoice-Small and SenseVoice-Large models can classify the audio event in the speech, including music, applause, and laughter. The SenseVoice-L can further predict the start and end position of the audio event, while the SenseVoice-Small can only predict what happened in the audio, with at most one event per utterance. SenseVoice-Small can detect more kinds of events, such as coughing, sneezing, breathing, and crying which could occur during human-machine interaction.

\begin{figure}[htb]
	\centering
	\begin{minipage}{0.44\linewidth}
		\vspace{4pt}
		\centerline{\includegraphics[width=\textwidth, trim={0.2cm 0.2cm 0.2cm 1.0cm}, clip]{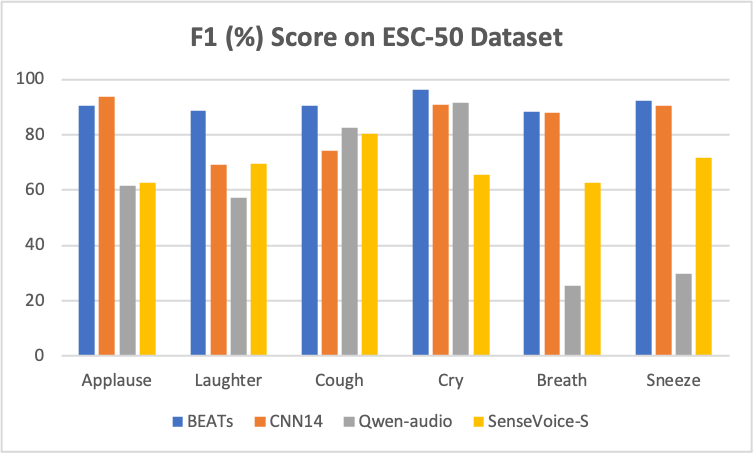}}
		\centerline{ESC-50}
	\end{minipage}
	\begin{minipage}{0.44\linewidth}
		\vspace{4pt}
		\centerline{\includegraphics[width=\textwidth, trim={0.2cm 0.2cm 0.2cm 1.0cm}, clip]{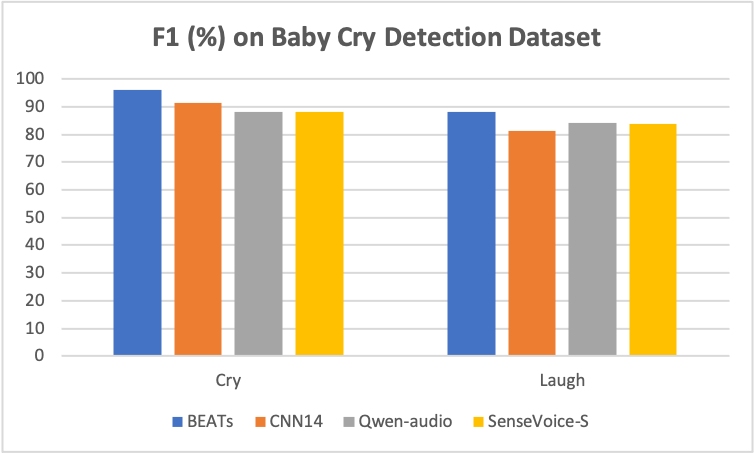}}
		\centerline{Baby Cry Detection}
	\end{minipage}
	
	\begin{minipage}{0.44\linewidth}
		\vspace{4pt}
		\centerline{\includegraphics[width=\textwidth, trim={0.2cm 0.2cm 0.2cm 1.0cm}, clip]{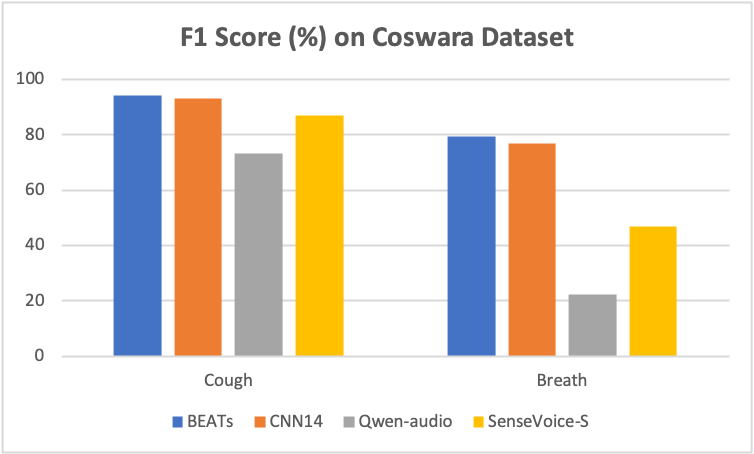}}
		\centerline{Coswara}
	\end{minipage}
	\begin{minipage}{0.44\linewidth}
		\vspace{4pt}
		\centerline{\includegraphics[width=\textwidth, trim={0.2cm 0.2cm 0.2cm 1.0cm}, clip]{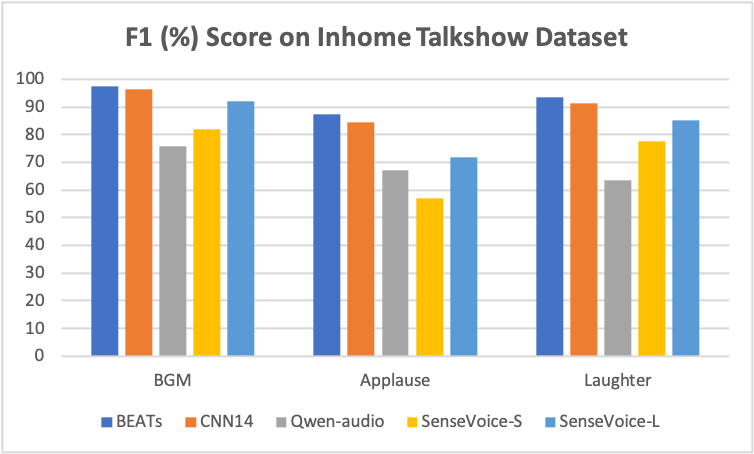}}
		\centerline{Inhome Talkshow}
	\end{minipage}
	\caption{F1(\%) Score comparison of the SenseVoice with the audio event detection models BEATS and PANNs on different audio event detection tasks.} \label{aed_results}
\end{figure}

We compare SenseVoice with the SOTA audio event detection models BEATs\citep{BEATs} and PANNs\citep{pann} on different tasks, including environment sound classification (ESC-50)\citep{esc50}, baby cry/laugh detection\footnote{\url{https://github.com/giulbia/baby_cry_detection/tree/master}}, coughing detection (Coswara)\citep{Coswara} \footnote{\url{https://github.com/iiscleap/Coswara-Data/tree/master}} and in-home talkshow event detection. As SenseVoice only predicts the event of our interest, which may not include event categories in other models, we use the F1 score on each event for evaluation. Qwen-audio is also evaluated for comparison.

We find that SenseVoice serves as a good audio event classification or detection model, though BEATs and PANNs may have better F1 scores, which may be attributed to two reasons. Firstly, BETAS and PANNs can modify the detection threshold to trade-off the accuracy and recall rate to obtain a higher F1 score, but threshold modification is much more difficult for SenseVoice and Qwen-Audio (An interesting discovery is that SenseVoice and Qwen-Audio always have a much higher accuracy than the recall rate, which could be more friendly for the human-machine interaction). Secondly, SenseVoice is trained with ASR data with AED pseudo labeling rather than AED-specific data. 

\subsection{Preserving Semantic Information by $S^3$ Tokenizer}
To assess the $\mathcal{S}^3$ tokenizer's ability to preserve semantic information, we compared the recognition performance of the quantizer-augmented SenseVoice-L against its original version and the Whisper-Large V3 model. 
The models underwent evaluation using the Common Voice zh-CN and en benchmarks, with the findings detailed in Table \ref{tab:tokenizer-performance}.

From the table, we can see that our $\mathcal{S}^3$ tokens demonstrate robust recognition performance in both the Chinese and English test sets. Notably, on the common\_voice\_zh-CN set, $\mathcal{S}^3$ tokens surpass the performance of the Whisper-Large V3 model, achieving a 4.14\% relative reduction in error rate. This suggests a substantial correlation between $\mathcal{S}^3$ tokens and semantic content.
It is worth noting that there is only a single codebook in the $\mathcal{S}^3$ tokenizer with a dictionary size of 4,096 entries.

\begin{table}[h]
	\centering
	\setlength\tabcolsep{6pt}
	\scalebox{1.0}{
		\begin{tabular}{l|cc|cc|cc}
			\toprule
			& \multicolumn{2}{c|}{Whisper-L-V3} & \multicolumn{2}{c|}{SenseVoice-L} & \multicolumn{2}{c}{$\mathcal{S}^3$ tokens} \\
			Test set & w/o lid & w/ lid & w/o lid & w/ lid & w/o lid & w/ lid \\\hline
			common\_voice\_zh-CN & 12.82 & 12.55 & 8.76 & 8.68 & 12.24 & 12.06 \\
			common\_voice\_en & 13.55 & 9.39 & 9.79 & 9.77 & 15.43 & 15.38 \\
			\bottomrule
	\end{tabular}}
	\vspace{0.1cm}
	\caption{The evaluation on $\mathcal{S}^3$ tokens' capability to preserve semantic information. We employ character and word error rates for zh-CN and en languages on the Common Voice benchmarks. Please note that the SenseVoice-L model in this table is an intermediate version, and is not identical to the one presented in Table \ref{tab:performance1}.}
	\label{tab:tokenizer-performance}
\end{table}

\subsection{Evaluation on Generation Quality of CosyVoice}
We evaluate the quality of CosyVoice's speech synthesis by examining content consistency and speaker similarity.
The ``test-clean'' subset of LibriTTS \citep{zen2019libritts} and the test set of AISHELL-3 \citep{DBLP:conf/interspeech/ShiBXZL21} are employed to construct evaluation set for English and Chinese, respectively. For each text in these sets, we randomly select a prompt speech. Content consistency was evaluated using Whisper-Large V3 \citep{DBLP:conf/icml/RadfordKXBMS23} for English and Paraformer~\citep{DBLP:conf/interspeech/GaoZ0Y22} for Chinese recognition. Speaker similarity was quantified by calculating the cosine similarity between speaker embeddings of the generated and prompt speeches, extracted using ERes2Net \citep{eres2net}.

Similar to other autoregressive language models, we employ a random sampling decoding strategy for our token LM and assessed the synthesis process using five different random seed values: 0, 7, 42, 123, and 1,337. The resultant evaluation metrics were averaged to determine the mean and standard deviation. Additionally, we conducted an ASR re-ranking to demonstrate potential performance improvements in offline mode.

Tables \ref{tab:res-libritts} and \ref{tab:res-aishell} present the results for English and Chinese, respectively. On the English dataset, CosyVoice attained human-level performance with similar content recognition and higher speaker similarity. ASR re-ranking notably enhanced content consistency, yielding a reduced word error rate (WER) of 1.51\%. CosyVoice outperformed ChatTTS in WER and the number of insertion and deletion errors, indicating superior content consistency. We did not assess speaker similarity for ChatTTS as it doesn't release voice cloning capabilities.

\begin{table}[h]
	\centering
	\setlength\tabcolsep{15pt}
	\scalebox{1.0}{
		\begin{tabular}{lccc}
			\toprule
			Model & WER (\%) & \#Ins.\&Del. & SS \\
			\midrule
			Original & 2.66 & 92 & 69.67 \\ 
			ChatTTS & 8.32 & 441 & - \\
			CosyVoice & 2.89$\pm$0.18 & 88.60$\pm$3.88 & 74.30$\pm$0.15 \\
			\ \ + 5$\times$ re-ranking & 1.51 & 47 & 74.30 \\
			\bottomrule
	\end{tabular}}
	\vspace{0.15cm}
	\caption{The comparison of original and CosyVoice generated speeches on the LibriTTS test-clean set in terms of word error rate (WER) and speaker similarity (SS). ``$\pm$'' joins the mean and standard deviation for each evaluation metric.}
	\label{tab:res-libritts}
\end{table}

\begin{table}[h]
	\centering
	\setlength\tabcolsep{15pt}
	\scalebox{1.0}{
		\begin{tabular}{lccc}
			\toprule
			Model & CER (\%) & \#Ins.\&Del. & SS \\
			\midrule
			Original & 2.52 & 25 & 74.15 \\ 
			ChatTTS & 3.87 & 111 & - \\
			CosyVoice & 3.82$\pm$0.24 & 24.4$\pm$2.24 & 81.58$\pm$0.16 \\
			\ \ + 5$\times$ re-ranking & 1.84 & 11 & 81.58 \\
			\bottomrule
	\end{tabular}}
	\vspace{0.15cm}
	\caption{The comparison of original and CosyVoice generated speeches on the AISHELL-3 test set in terms of character error rate (CER) and speaker similarity (SS). ``$\pm$'' joins the mean and standard deviation for each evaluation metric.}
	\label{tab:res-aishell}
\end{table}

As for the results on Chinese, the generated utterances of CosyVoice achieves a comparable CER as well as the errors of insertion and deletion compared with the original utterances. It seems that ChatTTS has a better generation ability on Chinese than English in terms of CER. Although ChatTTS and CosyVoice achieves a similar CER, ChatTTS produces more insertion and deletion errors, This is due to the problem of speaker leaking, where modal particles of another speaker is generated unexpectedly. On the contrary, CosyVoice doesn't suffer this problem with much less insertion and deletion errors. With ASR re-ranking, CosyVoice reached a remarkably low CER of 1.84\%. As seen with English, CosyVoice also exhibited greater speaker similarity than the original utterances, showcasing its effective voice-cloning proficiency.

\subsection{Evaluation on Emotion Controllability of CosyVoice}
To verify the emotion controllability, we use the public speech emotion recognition model emo2vec\footnote{\url{https://modelscope.cn/models/iic/emotion2vec_base_finetuned}} \citep{DBLP:journals/corr/abs-2312-15185}. We generate and evaluate 100 English utterances for each of the six emotions: happy, angry, sad, surprised, fearful, and disgusted. The content of the synthesized text is designed to match the target emotion. We then measure the accuracy of the predicted emotions from the synthesized speech for each emotion. 

Table \ref{tab:emo_acc} shows the comparison of emotion control accuracy between CosyVoice-base and CosyVoice-instruct. For CosyVoice-instruct, the input consists of content text accompanied by a speaking style instruction (e.g., ``Happy.$<$endofprompt$>$Content Text''). In contrast, CosyVoice-base only receives the content text as input. 
The results indicate that CosyVoice-instruct with emotional instructions demonstrates a significant improvement over both CosyVoice-base and CosyVoice-instruct without emotional instructions.

\begin{table}[h]
	\centering
	\scalebox{0.9}{
		\begin{tabular}{l c c c c c c}
			\toprule
			Model & Happy & Sad & Angry & Surprised & Fearful & Disgusted \\
			\midrule
			CosyVoice-base & 1.00$\pm$0.00 & 0.45$\pm$0.05 & 0.59$\pm$0.03 & 0.26$\pm$0.02 & 0.88$\pm$0.01 & 0.46$\pm$0.06  \\
			CosyVoice-instruct & 1.00$\pm$0.00 & 0.98$\pm$0.02 & 0.83$\pm$0.04 & 0.64$\pm$0.03 & 0.87$\pm$0.03 & 0.93$\pm$0.02  \\
			~~w/o instruction & 0.98$\pm$0.01 & 0.77$\pm$0.04 & 0.49$\pm$0.12 & 0.28$\pm$0.06 & 0.83$\pm$0.04 & 0.45$\pm$0.16 \\
			\bottomrule
	\end{tabular}}
	\vspace{0.15cm}
	\caption{Comparison of emotion control accuracy between CosyVoice-base and CosyVoice-instruct. ``$\pm$'' joins the mean and standard deviation for each evaluation metric.}
	\label{tab:emo_acc}
\end{table}

\subsection{CosyVoice as a Data Generator}
A straightforward application of CosyVoice is as a data generator to augment the training data of other tasks, such as ASR, speech-to-speech translation (S2ST). Taking the ASR task an example, we conduct an experiment on the Librispeech corpus to evaluate CosyVoice's capability in generating high-quality data. The experimental results are shown in Table \ref{tab:data-syn}, where ``Librispeech'' denotes the original 960-hour data. ``Syn on LS text'' and ``Syn on LS text'' denote the generated data with the text from Librispeech and MLS training sets, respectively. From the table, we can see that only training on the synthesized data, the ASR model can achieve a comparable result than the original Librispeech training set. Upon integration of them, a notable enhancement in recognition accuracy is observed. An interesting finding is that involving the synthesized data on the MLS text significantly improve the recognition performance. This may indicates that the text diversity is more critical for ASR task than the duration of speech itself. This improvement can be attributed to the varied linguistic content introduced by CosyVoice synthesized samples. The findings from our evaluation underscore the high quality of the samples generated by CosyVoice.

\begin{table}[h]
	\centering
	\setlength\tabcolsep{3pt}
	\scalebox{0.9}{
		\begin{tabular}{lcccc}
			\toprule
			Training Data & dev\_clean & dev\_other & test\_clean & test\_other \\
			\midrule
			Librispeech & 2.77 & 5.84 & 2.79 & 5.97 \\ 
			Syn on LS text & 2.79 & 6.37 & 3.00 & 6.59 \\ 
			Librispeech + Syn on LS text & 2.44 & 5.52 & 2.56 & 5.68 \\ 
			Librispeech + Syn on LS text $\times 2$ & 2.51 & 5.23 & 2.68 & 5.26 \\ 
			Librispeech + Syn on LS, MLS text & \textbf{1.93} & \textbf{4.43} & \textbf{2.04} & \textbf{4.53} \\
			\bottomrule
	\end{tabular}}
	\vspace{0.15cm}
	\caption{Evaluation on CosyVoice generation quality by treating it as a data generator. Word error rates (\%) on the human-uttered test sets are employed as the evaluation metrics.}
	\label{tab:data-syn}
\end{table}

\section{Applications}
The FunAudioLLM is an innovative framework designed to facilitate natural voice interactions between humans and large language models (LLMs). By integrating SenseVoice, CosyVoice, and LLMs, FunAudioLLM offers a variety of rich application demos, including speech-to-speech translation, emotional voice chat, interactive podcasts, and expressive audiobook narration. The demos are available at \url{https://fun-audio-llm.github.io}.

\begin{figure}[h]
\centering
\includegraphics[width=1.0\linewidth]{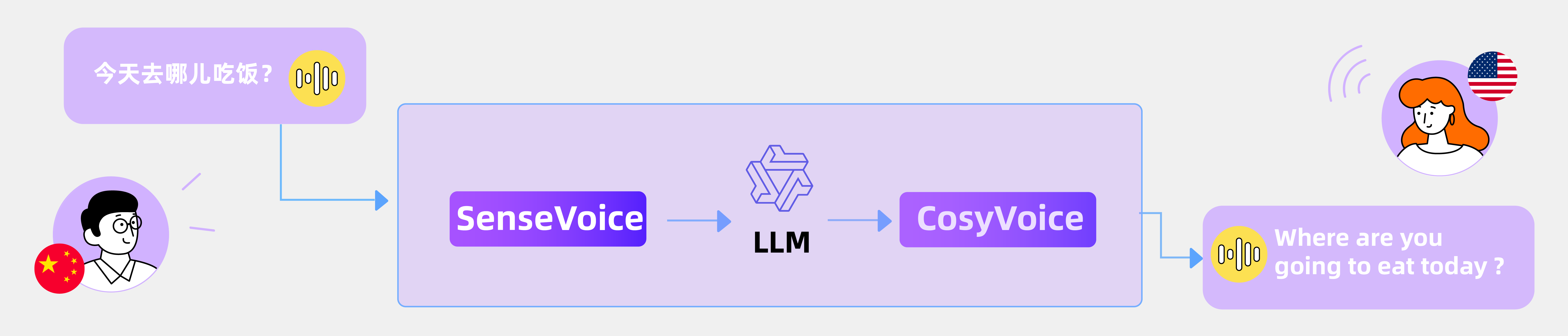}
\vspace{-7mm}
\caption{A diagram of Speech-to-Speech Translation.}
\label{fig:s2st}
\end{figure}

By combining SenseVoice, LLMs, and CosyVoice, we can effortlessly perform speech-to-speech translation (S2ST), as illustrated in Figure \ref{fig:s2st}. SenseVoice is used to recognize the input speech in its original language, the LLM translates the source language to the target language, and CosyVoice synthesizes the target speech with cross-lingual voice cloning. This allows users to speak in foreign languages using their own voice.

\begin{figure}[h]
\centering
\includegraphics[width=1.0\linewidth]{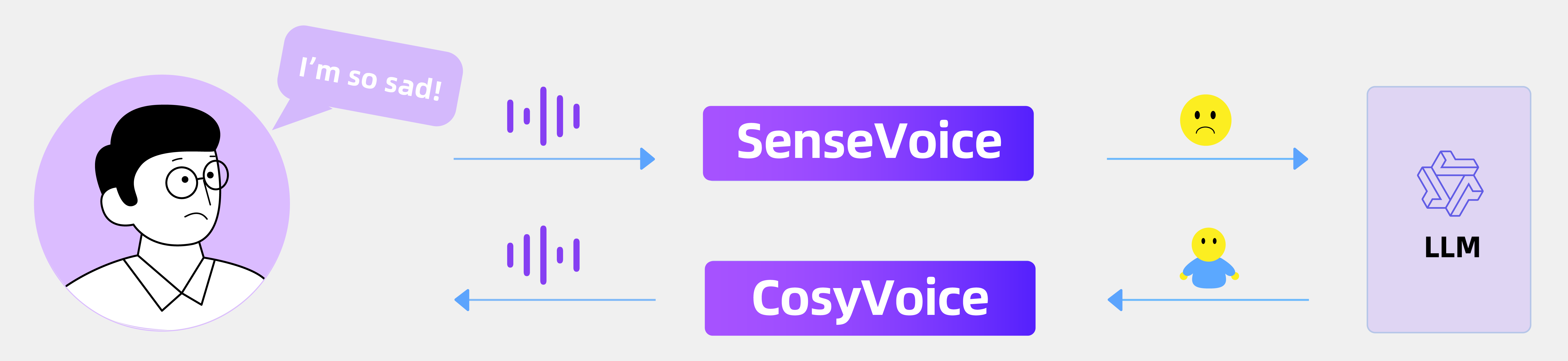}
\vspace{-7mm}
\caption{A diagram of Emotional Voice Chat.}
\label{fig:emotionalvoicechat}
\end{figure}

By integrating SenseVoice, LLMs, and CosyVoice, we can develop an Emotional Voice Chat application, as depicted in Figure \ref{fig:emotionalvoicechat}. SenseVoice recognizes the input speech and its emotion and audio event, the LLM generates the response content with a speaking style description, and CosyVoice produces emotional speech following the given speaking style description.

\begin{figure}[h]
\centering
\includegraphics[width=1.0\linewidth]{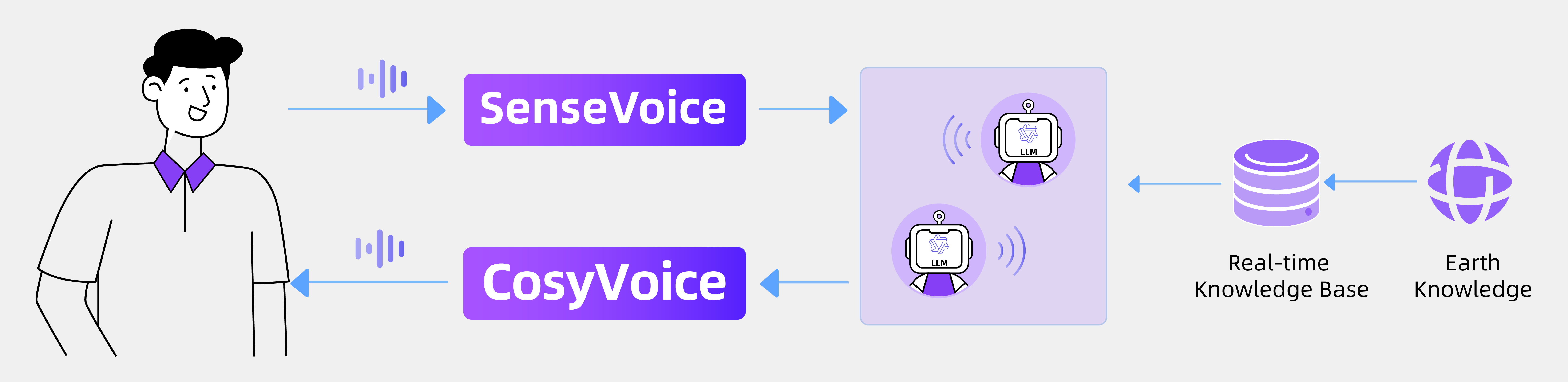}
\vspace{-7mm}
\caption{A diagram of Interactive Podcast.}
\label{fig:podcast}
\end{figure}

By leveraging SenseVoice, an LLM-based multi-agent system with real-time world knowledge, and CosyVoice, we can create an interactive podcast, as shown in Figure \ref{fig:podcast}. We can use an LLM plugin to fetch real-time daily knowledge, which a content-generation agent then transforms into a podcast script. The Multi-Agent system matches podcast roles, and CosyVoice synthesizes the voices. Users can also insert themselves into the podcast for interactive dialogues with the Multi-Agent system.

\begin{figure}[h]
\centering
\includegraphics[width=1.0\linewidth]{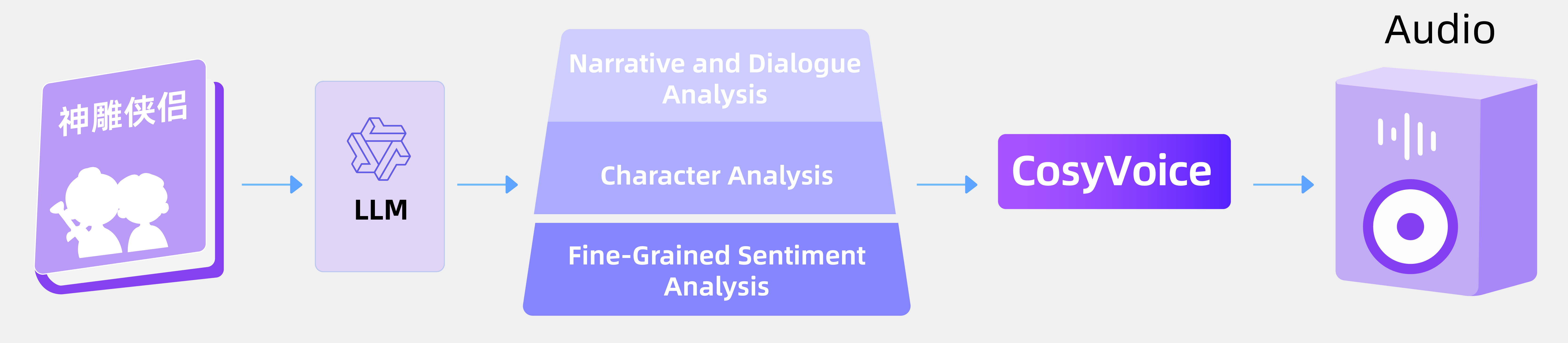}
\vspace{-7mm}
\caption{A diagram of Expressive Audiobook.}
\label{fig:audiobook}
\end{figure}

Through the analytical capabilities of LLMs to structure and identify emotions within books, and synthesizing this with CosyVoice, we achieve audiobooks with enhanced expressiveness, as illustrated in Figure \ref{fig:audiobook}. The LLM is used for narrative and dialogue analysis, character analysis, and fine-grained sentiment analysis, while CosyVoice synthesizes the speech with enhanced expressiveness.

\section{Limitations}
SenseVoice has certain limitations that need to be addressed. Firstly, the ASR performance generally remains much lower for under-resourced languages. Secondly, SenseVoice is not designed for streaming transcription. Therefore, future work may focus on developing streamable voice understanding models based on SenseVoice.

CosyVoice also has several limitations. Firstly, it supports a limited number of languages. While it can express emotions and speaking styles based on explicit instructions, it cannot infer the appropriate emotion or style based on the semantic content of the text. Additionally, CosyVoice does not perform well when tasked with singing. There's still room for improvement in achieving expressive emotional changes while maintaining the original timbre of the voice.

Another limitation is that the two innovative models within FunAudioLLM are not trained end-to-end with LLMs. This pipeline approach may introduce error propagation, which could affect overall performance.


\section{Authors (alphabetical order of family name)}
\begin{multicols}{3}
	\begin{itemize}[noitemsep]
		\item Keyu An
		\item Qian Chen
            \item Chong Deng
		\item Zhihao Du
		\item Changfeng Gao
		\item Zhifu Gao
		\item Yue Gu
		\item Ting He
		\item Hangrui Hu
		\item Kai Hu
            \item Shengpeng Ji
            \item Yabin Li
		\item Zerui Li
		\item Heng Lu
		\item Haoneng Luo
		\item Xiang Lv
            \item Bin Ma
		\item Ziyang Ma
            \item Chongjia Ni
            \item Changhe Song
            \item Jiaqi Shi
            \item Xian Shi
            \item Hao Wang
		\item Wen Wang
		\item Yuxuan Wang
		\item Zhangyu Xiao
		\item Zhijie Yan
		\item Yexin Yang
            \item Bin Zhang
		\item Qinglin Zhang
		\item Shiliang Zhang
		\item Nan Zhao
            \item Siqi Zheng
	\end{itemize}
\end{multicols}

\section{Acknowledgment}
We extend our heartfelt appreciation to the developers and contributors of the following open-source projects: FunASR, FunCodec, Whisper, ESPNet, WeNet, SLAM-LLM, Matcha-TTS, and Tortoise. Their innovative efforts and valuable code contributions have significantly inspired our work and facilitated our research. We are also grateful to numerous other projects not explicitly mentioned here, which have equally provided considerable assistance and played an instrumental role in the success of our endeavors.

\bibliographystyle{iclr2023_conference}
\bibliography{refs}

\appendix
\section{Auxiliary Results of SenseVoice on Common Voice.}
\begin{table}[ht]

\centering
\scalebox{1.0}{
\begin{tabular}{llcccc}
\toprule
& &  \multicolumn{2}{c}{Whisper-L-V3} & \multicolumn{2}{c}{SenseVoice-L} \\
Language & & w/o lid & w lid & w/o lid & w lid \\\hline
zh-CN & & 12.82 & 12.55 & 7.92 & 7.68 \\
\midrule
en & & 13.55 & 9.39 & 14.30 & 9.00 \\
\midrule
yue & & 40.42 & 10.51 & 7.08 & 6.78 \\
\midrule
ja & & 11.18 & 10.34 & 9.58 & 9.19 \\
\midrule
ko & & 5.59 & 5.59 & 5.23 & 5.21 \\
\midrule
fr & & 11.13 & 10.77 & 8.67 & 8.45 \\
\midrule
es & & 5.00 & 4.74 & 5.37 & 4.63 \\
\midrule
it & & 5.93 & 5.46 & 5.74 & 5.16 \\
\midrule
ru & & 6.16 & 5.67 & 6.60 & 5.23 \\
\midrule
id & & 8.98 & 7.22 & 12.80 & 6.97 \\
\midrule
th & & 9.73 & 5.80 & 4.36 & 4.12 \\
\midrule
de & & 6.06 & 5.70 & 6.94 & 6.57 \\
\midrule
ca & & 16.76 & 13.20 & 5.90 & 5.62 \\
\midrule
nl & & 5.51 & 4.28 & 6.65 & 5.23 \\
\midrule
pt & & 6.90 & 5.92 & 9.05 & 6.88 \\
\midrule
pl & & 7.26 & 5.95 & 10.01 & 7.47 \\
\midrule
cs & & 10.99 & 9.04 & 11.45 & 9.70 \\
\midrule
hi & & 46.17 & 16.88 & 48.85 & 10.06 \\
\midrule
tr & & 14.65 & 12.04 & 14.10 & 11.09 \\
\midrule
ro & & 13.43 & 10.84 & 18.21 & 12.01 \\
\midrule
hu & & 13.89 & 13.40 & 12.53 & 12.27 \\
\midrule
da & & 15.59 & 12.49 & 17.41 & 13.23 \\
\midrule
bg & & 17.05 & 14.24 & 18.56 & 13.25 \\
\midrule
mr & & 38.14 & 31.13 & 20.80 & 13.51 \\
\midrule
el & & 15.58 & 13.73 & 25.39 & 16.98 \\
\midrule
uk & & 15.89 & 11.60 & 12.43 & 20.75 \\
\midrule
az & & 36.32 & 25.21 & 72.65 & 28.63 \\
\midrule
sw & & 54.10 & 50.43 & 26.21 & 25.85 \\
\midrule
fa & & 37.44 & 34.86 & 32.40 & 40.33 \\
\midrule
bn & & 42.25 & 40.15 & 44.10 & 43.80 \\
\bottomrule
\end{tabular}}
\vspace{0.15cm}
\caption{Performance comparisons among different models with and without language id.}
\label{tab:performance}
\end{table}

\end{document}